%% file: main.tex
\documentclass[a4paper,twoside]{article}

\usepackage{epsfig}
\usepackage{subfigure}
\usepackage{calc}
\usepackage{amssymb}
\usepackage{amsmath}
\usepackage{amsfonts}
\usepackage{array}
\usepackage{graphicx}
\usepackage{multicol}
\usepackage[bookmarks=true,hidelinks=false]{hyperref}
\usepackage{balance}
\usepackage{xspace}
\newcommand{\ie}{i.e.\xspace}
\usepackage[T1]{fontenc}
\usepackage[utf8]{inputenc}
\usepackage{placeins}

\providecommand{\SetAlCapHSkip}[1]{}
\providecommand{\SetAlCapNameFnt}[1]{}
\usepackage{SCITEPRESS}
\usepackage{amsthm}

\usepackage{longtable}
\usepackage{booktabs}
\usepackage{multirow}
\usepackage{xcolor}
\usepackage{colortbl}
\usepackage{url}
\usepackage{pifont}
\usepackage{fontawesome5}
\usepackage{tcolorbox}
\tcbuselibrary{breakable, skins}
\usepackage{algorithm}
\usepackage{algorithmic}
\usepackage{float}
\usepackage{varwidth}
\usepackage{tikz}
\usepackage{enumitem}
\usepackage{tcolorbox}
\tcbuselibrary{theorems}

\usepackage{apalike}
\definecolor{sectionbg}{RGB}{220,220,220}
\definecolor{subsecbg}{RGB}{230,230,230}
\definecolor{thisworkcolor}{RGB}{0,90,160}
\definecolor{bronzecolor}{RGB}{205,127,50}
\definecolor{silvercolor}{RGB}{140,140,140}
\definecolor{goldcolor}{RGB}{212,175,55}
\definecolor{diamondcolor}{RGB}{0,185,235}

\newcommand{\tlaplus}{TLA$^{+}$\xspace}
\newcommand{\passx}{$\sf{pass@1}$}
\newcommand{\ours}{TLA-Prover}

\newcommand*\cibbox[1]{\tikz[baseline=(char.base)]{\node[shape=rectangle,fill=gray!30!white,text=black,inner sep=1.5pt] (char) {#1};}}

\newcommand{\BfPara}[1]{\vspace{2pt}\noindent\cibbox{{\bf{{#1.}}}}}

\makeatletter
\renewcommand\subsection{%
  \@startsection{subsection}{2}{\z@}%
    {-5pt plus -1pt minus -1pt}%
    {3pt plus 1pt minus 1pt}%
    {\large\bf\raggedright}}
\renewcommand\subsubsection{%
  \@startsection{subsubsection}{3}{\z@}%
    {-4pt plus -1pt minus -1pt}%
    {2pt plus 1pt minus 1pt}%
    {\normalsize\bf\raggedright}}
\makeatother

\hypersetup{
	colorlinks=true,
	urlcolor=black,
	linkcolor=blue!75!black,
	citecolor=red!45!black,
} 

\newtcbtheorem[]{observation}{\textbf{}}%
{colframe=white!95!black,fonttitle=\bfseries,coltitle=black, boxsep = 0pt, left = 3pt, right = 0pt, top = 0pt, bottom = 0pt, boxrule = 0pt, bottomrule = 0pt, toprule = 0pt}{th}{}

\newtheoremstyle{styledef}%
{\topsep}
{\topsep}
{}
{1pt}
{\bfseries}
{.}
{3pt}
{}
\theoremstyle{styledef}

\newcommand{\observationdef}[2]{
  \vspace{1ex}\par\noindent\tikzstyle{mybox} = [fill=gray!10,
   thick,rectangle,inner sep=4pt,path picture={\fill [white!40!black] ([xshift=-104pt]path picture bounding box.north) rectangle (path picture bounding box.south west);}]
  \begin{tikzpicture}
  \centering
   \node [mybox] (box){%
    \begin{minipage}[c]{0.96\linewidth}
    {#1}{ #2}\end{minipage}
   };
  \end{tikzpicture}
}

\begin{document}

\title{\ours{}: Verifiable \tlaplus{} Specification Synthesis via Preference-Optimized Low-Rank Adaptation}

\author{
\authorname{Eric Spencer, 
Arslan Bisharat\orcidAuthor{0009-0005-7815-4311},
Brian Ortiz\orcidAuthor{0009-0002-4223-8159},
Khushboo Bhadauria,
Mujtaba Nazari\orcidAuthor{0009−0003−4731−1789},
TaiNing Wang\orcidAuthor{0009-0007-1185-8183},
George K.\ Thiruvathukal\orcidAuthor{0000-0002-0452-5571},
Konstantin Läufer\orcidAuthor{0000-0002-7548-0876},
Mohammed Abuhamad\orcidAuthor{0000-0002-3368-6024}}
\affiliation{Department of Computer Science, Loyola University Chicago, Chicago, IL 60660, USA}
\email{\{espencer2,marslan,bortiz4,kbhadauria,mnazari,twang12,gthiruvathukal,klaufer,mabuhamad\}@luc.edu}
}

\abstract{
\tlaplus{} is a formal specification language for verifying distributed systems and safety-critical protocols.
Large language models (LLMs) frequently produce \tlaplus{} specifications that fail the TLC model checker for semantic reasons.
Across 25 LLMs, the best public baseline is 26.6\% syntactic parse and 8.6\% semantic model-check.
We present \ours{}, a 20-billion-parameter model for \tlaplus{} specification synthesis.
Training combines supervised fine-tuning (SFT) on verified examples with repair-based group-relative policy optimization (GRPO).
In the GRPO stage, the model learns to fix its own rejected specifications.
We also train a direct preference optimization (DPO) variant from the same SFT checkpoint as an ablation.
TLC provides the reward signal directly, with no learned reward model.
Each output falls into one of four tiers: Bronze (parses), Silver (no warnings), Gold (passes TLC), and Diamond.
To reach Diamond, the model's correctness property is automatically altered in a small way; TLC must then detect a violation.
If TLC still passes, the property was always-true and contributes nothing; the output fails Diamond.
\ours{} reaches 9/30 (\ie \passx$=30\%$) at both Gold and Diamond on a held-out 30-problem benchmark.
This is roughly $3.5\times$ the 8.6\% untuned baseline.
The DPO variant reaches 20\% at Diamond.
Gold and Diamond coincide at every checkpoint; this prevents the trivial-property failure mode.}
\onecolumn \maketitle \normalsize \setcounter{footnote}{0} \vfill

\input{sections/introduction}
\input{sections/relatedworks}
\input{sections/proposedmethods}
\input{sections/experiments}
\input{sections/conclusion}

\balance
\bibliographystyle{apalike}
{\small
\bibliography{custom}}

\end{document}

%% file: sections/introduction.tex
\section{Introduction}\label{sec:introduction}

\tlaplus{}~\cite{lamport2002specifying} (Temporal Logic of Actions) is a formal
specification language that lets engineers express safety and liveness
properties of concurrent and distributed systems at a level of rigor
that admits machine checking.
In practice, \tlaplus{} has been applied at industrial
scale for more than a decade. Amazon Web Services has reported its
use since 2011 on high-availability cloud services such as DynamoDB,
S3 and Elastic Block Store~\cite{Newcombe2015}; Microsoft has
doumented its application to the Cosmos DB distributed
database~\cite{cirstea_validating_2024}. The language combines
temporal logic, first-order logic and set theory in a syntactic
surface that mainstream code corpora rarely contain, which makes
correct \tlaplus{} generation a demanding target for
general-purpose language models.

It pairs with the TLC model
checker~\cite{yu1999tlc}, which exhaustively explores the reachable
state space and returns concrete counterexample traces on failure.
These properties make \tlaplus{} a compelling
target for verifiable code generation by large language models (LLMs).
The verifier is deterministic, the success criterion is binary and a
failing run is informative rather than opaque.

Despite this appeal, general-purpose LLMs perform poorly on
\tlaplus{} generation. The language is sparsely
represented in public pre-training corpora, its syntax is
idiosyncratic (temporal operators, \texttt{CHOOSE}, set comprehension) and
passing the parser is only the first hurdle. A syntactically valid
specification can still fail the model checker if the declared
invariants do not capture the protocol's real properties, or if the
\texttt{CONSTANT} bindings do not satisfy the type assumptions of
\texttt{Init}. Surface-level fluency in \tlaplus{}
syntax does not guarantee model-checkable output.

Bisharat et al.~\cite{bisharat2026formallm} make this gap concrete in their FormaLLM study,
reporting 26.6\% syntactic parse (measured by SANY, the \tlaplus{} parser) and 8.6\% TLC model-check averaged
across 25 open-weight LLMs at the best single prompting strategy.
Fewer than one in ten generation attempts by an untuned model produces
a specification the model checker accepts. FormaLLM is a diagnostic
study; it maps the baseline but does not close it and its authors
explicitly call for task-trained models as the natural next step. The
present work is a direct response to that call.

Training a model for formal verification introduces a reward-hacking risk
that does not appear in natural-language settings. A model that
discovers tautological invariants such as \texttt{TypeOK == TRUE}
will pass TLC on every prompt while conveying no useful property of
the protocol. To make this concrete: for a counter protocol with bound $N$,
the invariant \texttt{TypeOK ==} $count \in 0{..}N$ fails TLC when mutated to $count \in 1{..}N$, while
\texttt{TypeOK == TRUE} is insensitive to any such mutation.
We address this with a four-tier validation hierarchy
whose top tier requires that mutating the invariant predicate causes
TLC to find a counterexample. This prevents the tautology issue by
construction.
This paper presents \ours{}, built on
\texttt{gpt-oss-20b}~\cite{openai2025gptoss} (20B parameters). Our
best checkpoint reaches 9/30 (30\%) at the TLC-diamond tier on a
held-out 30-problem benchmark, roughly $3.5\times$ the 
averaged-baseline TLC pass-rate of 8.6\%.

\BfPara{Contributions} We make three contributions.

\begin{itemize}[leftmargin=*]
\item We introduce \ours{} with a {four-tier validation hierarchy} (bronze, silver, gold,
diamond) whose top tier uses mutation testing to rule out tautological
invariants and addresses the reward-hacking risk that makes TLC an
otherwise exploitable reward signal.

\item We provide insights to \ours{} during the {two-stage training pipeline} that combines SFT on a
diamond-tier-curated corpus with repair-based GRPO using TLC as the
verifier reward, plus a DPO ablation for comparison. This takes us
from 0/30 to 9/30 diamond on the held-out benchmark.

\item We provide an {anti-reward-hacking analysis} which shows that gold and
diamond pass-rates coincide at every checkpoint on our held-out suite.
This empirically rules out the `\textit{always-true invariant}' failure mode.
\end{itemize}

To our knowledge, this is the first \tlaplus{}
targeted training study to substantially exceed the FormaLLM baselines.
The model, data and evaluation pipeline will be released on
acceptance. 

The underlying thesis of this paper is direct. TLC pass alone is an exploitable reward signal. Mutation-sensitive testing is what separates meaningful specifications from trivial ones. Verifier-guided training without anti-reward-hacking validation does not solve the problem.

\BfPara{Organization} \autoref{sec:relatedworks} covers related work.
\autoref{sec:methods} describes the training pipeline and evaluation harness.
\autoref{sec:experiments} reports results and six observations.
\autoref{sec:conclusion} concludes.

%% file: sections/relatedworks.tex
\section{Related Work}
\label{sec:relatedworks}

\BfPara{LLMs for Formal-Method} The most relevant prior work trains LLMs to produce artifacts
accepted by a deterministic formal checker. DeepSeek-Prover~\cite{xin2024deepseek}
generates 8M synthetic Lean~\cite{lean4} theorems and uses them to train a 7B
model, achieving state-of-the-art results on miniF2F and ProofNet
entirely through verifier-graded synthetic data. Baldur~\cite{first2023baldur}
targets whole-proof generation in Isabelle~\cite{paulson2000isabelle700theoremprovers} with a repair loop. When
the prover rejects a generated proof, the error message is fed back
into the model and a fresh attempt is solicited. Both works share the
key design choice we adopt: the deterministic verifier acts directly
as the reward signal, avoiding the instability of a learned reward model.
Neither targets \tlaplus{}. The reward-hacking surface also differs in kind.
In Lean or Isabelle, a shortcut proof still requires a non-trivial goal statement.
For TLC, the shortcut is an always-true invariant embedded in the specification
itself, reachable on every prompt without any such precondition.
Baldur's repair loop and our repair-based GRPO share the same structure.
In both, the model reads a rejection message and produces a revised attempt.
The difference is the tautology risk: Isabelle proof repair has no tautological escape route.
Specification repair does and the diamond tier closes it.

\BfPara{\tlaplus{} Generation and Synthesis}
Recent study \cite{bisharat2026formallm} provides FormaLLM, the first
systematic evaluation of LLM-based \tlaplus{} generation,
testing 30 models across eight families under four prompting
strategies. Their headline finding of 26.6\% SANY parse and 8.6\%
TLC model-check averaged across 25 open-weight models establishes
the untuned-LLM floor our work targets. FormaLLM is an evaluation
study; no task-specific training is performed and the authors explicitly
identify task-trained models as the natural next step. This
paper answers their call directly. We use the same benchmark
distribution and report what a targeted training pipeline adds on
top of the untuned baseline.

\BfPara{Preference Optimization} Direct preference optimization (DPO)~\cite{rafailov2023direct} fits
a preference objective on pairs of ranked model outputs without
requiring an explicit reward model and has been applied to
formal-method tasks to steer generation away from common syntax
errors. Group relative policy optimization
(GRPO)~\cite{shao2024deepseekmath} normalizes advantages within a
group of samples from the same prompt; it reduces variance without a
critic network and is the RL algorithm of choice for verifier-graded
verifier-graded training in DeepSeek-Prover and related work. However, standard
GRPO stops learning when rewards are sparse. When all $K$
samples from a prompt fail TLC, every output looks equally bad;
the group-relative advantage is zero and no gradient flows. This failure directly motivates the
repair-based GRPO shaping we describe in \autoref{sec:methods}.

\BfPara{Reward Hacking in LLM Verification}
The reward-hacking literature ~\cite{skalse2022defining,gao2023scaling}
documents the failure mode where a model optimizes a proxy signal in
a way that diverges from the intended objective. For
\tlaplus{} with TLC as the reward, the canonical hack is
the always-true invariant. A property such as \texttt{TypeOK == TRUE}
satisfies TLC on every reachable state while conveying nothing useful
about the protocol. Every \tlaplus{} prompt admits a specification
whose invariant is always satisfied, so an optimizer targeting TLC pass
alone reaches this shortcut on every prompt. A DPO or GRPO optimizer
targeting TLC pass alone would readily learn this hack. Our diamond tier
(\autoref{sec:four_tier}) closes that gap by requiring the invariant
to be sensitive to mutations.
DeepSeek-Prover, Baldur and \ours{} are all instances of the same pattern.
In each, a deterministic verifier replaces a learned reward model.

\BfPara{Industrial Formal Methods} Industrial adoption of \tlaplus{} in distributed systems is well established.
Amazon Web Services reported \tlaplus{} use in DynamoDB, S3 and Elastic Block Store before 2011~\cite{Newcombe2015}.
Microsoft applied it to the Cosmos DB distributed database~\cite{cirstea_validating_2024}.
TLA+ specifications are written by engineers.
Errors caught by TLC during design prevent expensive bugs in deployment.
This industrial workflow creates the concrete problem \ours{} targets.
A model that produces a checkable first-draft specification lowers the barrier for teams new to \tlaplus{}.

%% file: sections/proposedmethods.tex
\section{Proposed Methods}
\label{sec:methods}

\BfPara{Four-Tier Validation Hierarchy}
\label{sec:four_tier}
Every model output is graded by a deterministic external pipeline
into exactly one of four tiers, where each tier requires all
predicates of the previous tier plus one additional check.

\begin{itemize}[leftmargin=*]
\item \textbf{Bronze.} SANY rejects the output; the
\tlaplus{} parser failed and the output is not a valid
specification.

\item \textbf{Silver.} SANY passes but TLC does not; the model
checker fails to load, times out, or rejects the specification before
exhausting the state space.

\item \textbf{Gold.} Both SANY and TLC pass; the model checker
exhausts the reachable state space within budget and all declared
invariants hold at every state.

\item \textbf{Diamond.} Gold and the invariant is mutation-sensitive.
We apply an automated AST-level mutation to the invariant predicate
and rerun TLC on the mutated specification; the mutation must cause a
TLC counterexample. Mutations include negating a top-level conjunct,
flipping $=$ to $\neq$ or $\leq$ to $\geq$, or swapping a named
constant for a random alternative from the same syntactic class.
Tautological invariants such as \texttt{TypeOK == TRUE} fail this
check by construction, because no mutation of \texttt{TRUE} produces
a state TLC would reject.
\end{itemize}

The bronze-to-gold ladder is the standard compile-and-check pipeline;
diamond adds one mutation-test step whose sole job is to deny credit
for trivial invariants. \autoref{fig:pipeline} shows the full
validation flow and \autoref{sec:experiments} reports how often
gold-tier outputs fail the diamond check empirically.

\input{figures/fig_pipeline}

\BfPara{Training Corpus}
We build the training corpus from two source pools and curate it
strictly to diamond tier, so that only specifications that are both
model-checkable and mutation-sensitive enter the final SFT corpus.

\BfPara{Seed Data} We bootstrap from the \tlaplus{}
Foundation specification distribution, the same 205-specification
corpus used by Bisharat et al.~\cite{bisharat2026formallm}, augmented
with the \texttt{tlaplus/examples} repository. That repository
contains 316 modules, of which 129 are TLC-targeted, 34 are
targeted at the TLA+ Proof System (TLAPS, a separate deductive verifier),
and the remainder are utility libraries or research drafts.

\BfPara{Curation Pipeline} A trained generator produces candidate
specifications from natural-language problem descriptions and each
candidate is graded by the four-tier pipeline. A separate frozen
LLM judge rates each candidate on six criteria: correct parsing,
successful model-checking, non-trivial invariants, no unused variables,
descriptive operator names and clean formatting. Only diamond-tier
survivors enter the SFT corpus, with diamond rows upsampled at
$2\times$, for a final corpus of 1{,}053 rows after deduplication.

\BfPara{Eval Set Carving} We hold out a 30-problem benchmark drawn
from the same descriptor distribution as training but with disjoint
module names. To measure judge agreement, we manually re-grade 60
random rows and observe 93\% agreement between the LLM-judge and a
human auditor; disagreements concentrate on borderline silver/gold
cases where TLC times out rather than passes.

\BfPara{Training Pipeline}
The production pipeline runs in two stages. Stage 1 applies
supervised fine-tuning (SFT) on the curated corpus to teach the model the
structural vocabulary of \tlaplus{}. Stage 2 applies
repair-based GRPO with TLC as the verifier reward to push the model
toward model-checkable and mutation-sensitive output. We also train a
DPO checkpoint from the same SFT base as a same-stage ablation; it
under-performs repair-GRPO on diamond pass-rate and is not part of
the final pipeline.

\BfPara{Stage 1. SFT} We train \texttt{gpt-oss-20b} with a LoRA
adapter (rank 8, alpha 16, dropout 0) applied to the attention
projections q/k/v/o of every transformer block. The MoE expert
weights are not targeted. Each block's 32 expert FFNs are stored as a
3D packed tensor that standard LoRA cannot decompose and the top-k
router is not a Linear module either, so attention is the only
trainable surface available without a custom adapter implementation.
The base model is dequantized from MXFP4 to BF16 for training. We
use AdamW with $\textit{lr}=2{\times}10^{-5}$ and a cosine schedule,
sequence length 3072, per-device batch size 1 and
gradient-accumulation 8. Training runs for 1 epoch on 2$\times$
NVIDIA RTX 8000 GPUs (48 GB each) via naive pipeline parallelism,
with a wall-time of 4--5 hours.

\BfPara{Stage 2. Repair-Based GRPO} Standard GRPO stops learning when rewards are sparse.
When all $K$ samples from a prompt fail TLC, every output looks equally bad;
the group-relative advantage is zero and no gradient flows. We address
this with repair-based shaping, where the model receives a broken
specification harvested from earlier generation attempts and is asked to repair it.
These self-repair trajectories are iterative runs in which the model
fixes its own prior outputs using TLC error messages as feedback. Trajectories are filtered to the
learnable middle ($0.10 \le$ before-score $\le 0.80$); this gives 42
broken/fixed pairs from a pool of 133 collected trajectories.
Common failure categories in the broken pool include random SANY-level
syntax violations, TLC-violating invariant replacements and
operator-name typos; the
model is asked to repair the broken spec to the highest tier it can
reach. The reward is a continuous improvement score from input tier to
output tier, weighted 0.1 (bronze to silver), 0.3
(silver to gold) and 0.6 (gold to diamond). This dense reward
produces non-zero gradient on every step where any sample improves.
We train for 160 steps with $K{=}4$ generations per prompt,
$\beta_{\text{KL}}{=}0.02$ and $T{=}0.5$. The optimizer is AdamW
with $lr{=}3{\times}10^{-6}$ and a linear-decay schedule.
Training completes in approximately 1.7 hours on the same
2$\times$ RTX~8000 hardware.
~\autoref{alg:repair_grpo} gives the full procedure.

\begin{algorithm}[!t]
\caption{Repair-Based GRPO}
\label{alg:repair_grpo}
{\small
\begin{algorithmic}[1]
\REQUIRE SFT checkpoint $\pi_\theta$; pool $\mathcal{B} = \{(\tilde{p}, p^{\star})\}$;
         grader \textsc{Grade}; weights $w_{\text{bs}}{=}0.1$, $w_{\text{sg}}{=}0.3$,
         $w_{\text{gd}}{=}0.6$; $N{=}160$, $K{=}4$, $\beta_{\text{KL}}{=}0.02$, $T{=}0.5$
\ENSURE  Repair-GRPO checkpoint $\pi_\theta$
\STATE Let $\mathcal{P} \leftarrow \{(\text{br},\text{si}),\,(\text{si},\text{go}),\,(\text{go},\text{di})\}$
\FOR{$t = 1$ \TO $N$}
  \STATE $(\tilde{p}, p^{\star}) \sim \mathcal{B}$
  \STATE $\tau_{\text{in}} \leftarrow \textsc{Grade}(\tilde{p})$
  \FOR{$k = 1$ \TO $K$}
    \STATE $s_k \sim \pi_\theta(\cdot \mid \tilde{p},\, T)$
    \STATE $\tau_k \leftarrow \textsc{Grade}(s_k)$
    \STATE $r_k \leftarrow \sum_{(a,b)\,\in\,\mathcal{P}}
           w_{ab}\cdot\mathbf{1}[\tau_{\text{in}} < b \leq \tau_k]$
  \ENDFOR
  \STATE $\bar{r} \leftarrow \tfrac{1}{K}\sum_k r_k$;\quad
         $\sigma_r \leftarrow \mathrm{std}(\{r_k\})$
  \STATE $A_k \leftarrow (r_k - \bar{r})\,/\,(\sigma_r + \epsilon)$
  \STATE $\mathcal{L} \leftarrow -\tfrac{1}{K}\sum_k A_k \log\pi_\theta(s_k \mid \tilde{p})$
  \STATE $\phantom{\mathcal{L} \leftarrow} + \;\beta_{\text{KL}}\,D_{\mathrm{KL}}(\pi_\theta \,\|\, \pi_{\text{SFT}})$
  \STATE Update $\theta$ by gradient step on $\mathcal{L}$
\ENDFOR
\RETURN $\pi_\theta$
\end{algorithmic}
}
\end{algorithm}

\BfPara{Stage 2-alt. DPO (ablation)} We sample 4 outputs per
training prompt from the SFT checkpoint and label each as winner or
loser by tier. A diamond-tier output beats any non-diamond, a
gold-tier output beats any non-gold and so on, with ties broken
arbitrarily. We use the standard DPO
objective~\cite{rafailov2023direct} with $\beta{=}0.1$ and the same
LoRA architecture as Stage 1. Training runs for 1 epoch on 343 preference pairs, sampled from a
pre-curation pool of candidate specifications that spans the full
descriptor distribution so that the SFT baseline's distributional
coverage is preserved.
The DPO checkpoint under-performs
repair-GRPO on diamond (see~\autoref{tab:per_stage}) and is not
used in the final pipeline.

\BfPara{Evaluation}
All evaluations run on the 30-problem held-out suite using the
four-tier validation pipeline. We report three inference strategies
to separate the contribution of the model from the contribution of
the inference procedure.

\BfPara{Inference Strategy} Headline numbers use single-shot greedy
decoding ($T{=}0$, $K{=}1$, no self-correction). We additionally
report best-of-$K$ ($T{=}0.5$, $K{=}8$) and the greedy $\cup$
best-of-K union as an upper-bound estimate of what the model can
produce given multiple attempts.

\BfPara{TLC Configuration} Each held-out spec is paired with a
\texttt{.cfg} file that fixes \texttt{CONSTANT}s at small finite
sizes (typically $N{=}3$ for $N$-process systems). TLC runs with a
60-second wall-clock budget; timed-out specs are graded silver unless
the partial state space already contains a violation, in which case
that violation counts as the grading signal.

\BfPara{Mutation Test Implementation} The mutation pass runs after a
gold-tier verdict. We select one of three AST mutation operators
uniformly at random. The first negates a top-level conjunct; the
second flips a relational operator ($=$ to $\neq$, $\leq$ to $\geq$);
the third replaces a named constant with a random alternative from
the same syntactic class. TLC is rerun on the mutated specification.
A counterexample means the original invariant passes the diamond
check; acceptance means it fails.

%% file: figures/fig_pipeline.tex
\begin{figure*}[!t]
\centering
\includegraphics[width=\textwidth, trim={0.6cm 0cm 1.3cm 0cm}, clip]{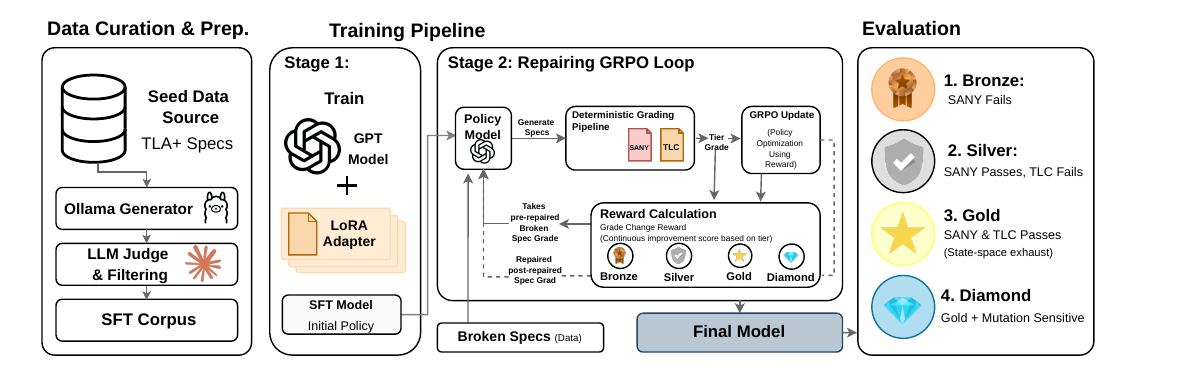}
\caption{End-to-end pipeline for verifiable \tlaplus{} specification synthesis and repair using Preference-Optimized Low-Rank Adaptation (LoRA).}
\label{fig:pipeline}
\end{figure*}

%% file: sections/experiments.tex
\section{Experiments and Results}
\label{sec:experiments}

\BfPara{Held-Out 30-Problem Benchmark}
We test our final checkpoint on a 30-problem held-out suite using
single-shot greedy decoding and the four-tier validation pipeline
throughout. The FormaLLM evaluation~\cite{bisharat2026formallm}
provides the published baseline. Pass-rates averaged across 25
open-weight LLMs on 26 specifications at the best single prompting
strategy, for a total of 650 runs per reported cell.~\autoref{tab:headline_results}
presents the comparison. The SFT-only row reports the best checkpoint
selected from six iterative SFT runs on the diamond-only corpus; internal
version labels are omitted as they carry no meaning for the reader.

\input{tables/tab_headline_results}

Our final checkpoint reaches 9/30 gold and 9/30 diamond on the
held-out 30-problem suite, roughly $3.5\times$ the FormaLLM
averaged open-weight-LLM TLC baseline of 8.6\%.
Because the held-out suite contains only 30 problems, each problem
swapped in or out moves the headline by 3.3 percentage points;
results should be interpreted with this variance in mind.

\input{tables/tab_per_stage}
\BfPara{Diamond-Gold Coincidence}
\autoref{tab:per_stage} shows that gold and diamond rates coincide
at every stage on the production path. The only gap appears at the
DPO ablation checkpoint, where one spec passed TLC but its invariant
was mutation-equivalent to \texttt{TRUE}. This is precisely the
failure mode the diamond tier was designed to catch and it is absent
from the final repair-GRPO checkpoint. Its absence provides strong empirical
evidence that the pipeline does not exploit TLC through the
trivial-tautology direction.

The 10-point gap between DPO and repair-GRPO on diamond (20\% vs
30\%) reflects the different quality of training signal. DPO learns
from static preference pairs sampled offline from the SFT checkpoint,
which limits signal diversity as the model improves. Repair-GRPO
receives a continuous dense reward on every step where any repaired
sample improves in tier; the result is richer gradient signal throughout
training. Subtler reward hacks remain possible; for example, a model could write
non-trivial but meaningless invariants that still pass TLC.
The simple always-true invariant hack is ruled out.~\autoref{fig:training_progression} shows the same progression visually.

\input{figures/fig_training_progression}

\BfPara{Per-Problem Breakdown}
The 30 held-out problems span six application domains. Gold and
diamond wins cluster in two categories. The first is invariant-style specifications
where a simple \texttt{TypeOK} state predicate suffices and problems
where both the descriptor and the canonical solution are small in
scope. Problems that resist solving share a common shape; they involve
multi-round protocols with phase transitions whose correctness
invariants require case-analysis over message types, such as Byzantine
broadcast and 3-phase commit.~\autoref{tab:domain_breakdown}
gives the per-domain breakdown.

\input{tables/tab_domain_breakdown}

\autoref{fig:output_example} illustrates the qualitative
difference between a diamond-tier output and a common failure. The
diamond output declares a non-trivial \texttt{TypeOK} invariant that
constrains the state space; mutating it causes TLC to find a
counterexample. The failing output either declares \texttt{TypeOK ==
TRUE} or generates a syntactically valid but semantically incoherent
\texttt{Init} predicate.

\input{figures/fig_output_example}

\BfPara{Curation Tier vs. Final Pass-Rate}
To measure the value of strict curation, we rebuild the SFT corpus
three times with different minimum tier cutoffs and re-train Stage 1
from scratch each time.~\autoref{tab:ablation_tier} shows the results.

\input{tables/tab_ablation_tier}

Despite being the smallest corpus, diamond-only training data produces
the best Stage-1 checkpoint. The result reinforces the dataset
quality-over-quantity principle and gives the diamond tier a dual
role. It is both the anti-reward-hacking criterion and the primary
curation filter for the training corpus.

\BfPara{Best-of-K and Self-Correction}
We test two augmented inference strategies on the final checkpoint
to separate the model's capability from the inference procedure.~\autoref{tab:augmented_inference} summarises the results.

\input{tables/tab_augmented_inference}

Best-of-8 sampling lifts the diamond rate by two problems and the
greedy-plus-best-of-8 union ceiling of 43\% indicates that the model
can produce correct solutions for 13 of 30 problems given sufficient
attempts. The 3-shot TLC-feedback repair pattern, however, produces no
improvement. On rows where greedy fails, the model reproduces the same
wrong proof structure across all retries even with the error message
in context. The pattern points to a knowledge gap rather than
a random sampling effect.

\BfPara{Key Observations Across Six Iterations}
\label{sec:discussion}
The results above represent the final state of a pipeline that passed
through six SFT iterations and two abandoned GRPO recipes. The
intermediate failures are instructive and we report them here in the
spirit of negative-result publishing.

\observationdef{%
\begin{observation*}{}{}
{\textbf{Observation 1.}} Corpus Labels Lie.
\end{observation*}
}

Our second-iteration corpus was labeled ``TLAPS-targeted'' and
contained 3{,}271 rows, but only 66 of those rows were genuinely
TLAPS content; the rest were spec-generation material co-located in
directories that happened to contain proof files. The unintentional
98\% spec / 2\% proof composition meant the model trained on this
corpus learned no prover capability.
The root cause was directory-level labeling: every file in a directory associated with proofs inherited the label without any content check.
A TLA+ repository commonly stores proof files and spec files side by side, so the label propagated to unrelated files.
Content-level auditing is the fix; file paths and directory names are not a substitute for inspecting row contents directly.

\observationdef{%
\begin{observation*}{}{}
{\textbf{Observation 2.}} Eval-set Size Matters for Variance.
\end{observation*}
}

An earlier evaluation harness used a 6-row holdout, which gives a
standard error of roughly 20 percentage points on the reported
pass-rate with $K=4$ greedy decodes. At that resolution we could not
distinguish genuine improvement from sampling noise.
The arithmetic is direct: with $n{=}6$ binary pass/fail outcomes, the binomial standard error is $\sqrt{p(1-p)/n} \approx 0.20$ at $p{=}0.5$.
A single problem flipping from fail to pass moves the headline by 16.7 percentage points.
A lucky random seed is statistically indistinguishable from a meaningful model improvement at that scale.
We carved the 30-row diamond suite specifically to bring the per-problem resolution to 3.3 percentage points.
A 100-problem holdout would reduce the standard error below 5 percentage points; we treat that as the minimum for domain-level analysis.

\observationdef{%
\begin{observation*}{}{}
{\textbf{Observation 3.}} GRPO Needs Reward Variance; Truncation Kills It.
\end{observation*}
}

Standard GRPO with $K=4$ and a 768-token completion budget hit a hard
ceiling because every sampled completion clipped at the budget cap and
produced the same floor reward, driving the group-relative advantage
to zero. Per-step GRPO with a 384-token budget surfaced the same
failure in a different shape.
The mechanism is precise: GRPO normalizes rewards as $A_k = (r_k - \bar{r})\,/\,(\sigma_r + \epsilon)$.
When all $K$ completions clip at the token budget, every output receives an identical bronze-tier reward.
The reward standard deviation $\sigma_r$ collapses to zero and the normalized advantage is undefined on every training step.
The model cannot improve when no sample in the group is better than any other.
Repair-based GRPO solves this structurally: the input is always a broken specification, so repair attempts that reach a higher tier produce non-zero variance by construction.
Monitor reward variance per step; if $\sigma_r \approx 0$ persists for more than a few steps, training has stalled regardless of what the loss curve reports.

\observationdef{%
\begin{observation*}{}{}
{\textbf{Observation 4.}} Stacking SFT Regresses.
\end{observation*}
}

Stacking incremental SFT runs on a prior LoRA checkpoint degraded
performance monotonically. The diamond pass-rate moved 4/30, 1/30,
0/30, 1/30 across four iterations.
The root cause is adapter drift: each new SFT step updates the LoRA weights relative to the current checkpoint, pulling them further from the frozen base model's parameter prior.
After two or three stacked runs, the effective update is a composition of gradients from different data snapshots.
The compounding drift corrupts syntactic fluency and semantic precision simultaneously; the model begins generating structurally plausible but type-incorrect specifications.
Restarting from the frozen base model at each major corpus revision was the fix.
Incremental SFT is acceptable only when the corpus tier distribution and domain mix are stable between runs.

\observationdef{%
\begin{observation*}{}{}
{\textbf{Observation 5.}} One Spec Template, Memorized.
\end{observation*}
}

The majority of diamond-tier outputs share the same skeletal structure,
declaring a TypeOK invariant as a state predicate and building a
simple Init/Next around small finite-state actions. Protocols that
require quantifier manipulation or case-analysis over message types
consistently produce malformed specifications.
The training corpus is dominated by invariant-style specifications with simple state spaces.
The model learned a high-reward template: declare TypeOK as a set-membership predicate, write a small Init, write a stuttering-step Next.
This template reliably reaches diamond tier on simple domains and therefore scores well throughout training.
Harder protocols that require \texttt{CHOOSE} expressions, recursive operators, or quantifier-bounded message sets receive no gradient signal because the model never produces a correct example of them.
The per-domain data in \autoref{tab:domain_breakdown} reflects this directly: mutual exclusion, scheduling, and workflows each score 2/3 diamond; consensus, data structures, and memory/caches each score 1/3; communication, concurrency, transactions, and puzzles each score 0/3.
Upsampling harder domains during SFT, even at silver tier, may force the model to build a wider syntactic vocabulary.

\observationdef{%
\begin{observation*}{}{}
{\textbf{Observation 6.}} Comments in Training Targets Leak into Outputs.
\end{observation*}
}

Earlier iterations included block-comment preamble inside assistant
targets and the model learned to emit verbose commentary before the
specification body. With a fixed 512-token completion budget this
caused the specification body to be truncated and stripping comments
from assistant targets recovered 5 percentage points of SANY parse rate.
The mechanism is instruction-tuning style imitation: the model treats the assistant-turn format as a style template.
When assistant targets consistently open with comment preambles, the model learns this as the expected output prefix.
Comment tokens consumed budget before the TLA+ module header appeared, leaving many specifications truncated mid-body.
Strip all comments from assistant targets, not just most of them.
Partial stripping leaves residual patterns that the model still learns.
The specification body must begin at the first token of the assistant turn; this applies to any verifier-graded setting where output must be parseable from its very first character.

\BfPara{Practitioner Guidance}
The following guidance applies to any team using \ours{} or a similar verifier-guided model.

\begin{itemize}[leftmargin=*]
\item \textbf{Do not treat TLC pass as sufficient.} Always run mutation testing. A gold-tier output without mutation sensitivity may still be vacuous.

\item \textbf{Use greedy decoding for reproducible evaluation.} Use best-of-$K$ sampling ($K \geq 4$) when the goal is a single usable candidate specification.

\item \textbf{Require human review.} Generated specifications are checkable first drafts, not verified designs.

\item \textbf{Expect domain variation.} The model performs best on small invariant-style specifications. Multi-round protocols with message-type case analysis are currently out of scope.

\item \textbf{Prefer diamond-tier training data.} Smaller corpora curated to diamond tier outperform larger silver- or gold-tier corpora (see \autoref{tab:ablation_tier}).
\end{itemize}

%% file: tables/tab_headline_results.tex
\begin{table*}[!t]
\centering
\small
\caption{Single-shot greedy results on the 30-problem held-out
suite. The FormaLLM row is the averaged open-weight-LLM pass-rate at
the best single prompting strategy from~\cite{bisharat2026formallm}
(25 models $\times$ 26 specs = 650 runs per cell). \ours{} rows are
single trained checkpoints tested at $T{=}0$, $K{=}1$. The DPO
row is a same-base ablation trained from the SFT checkpoint; it is not
part of the final pipeline (see \S\ref{sec:methods}).
\textit{Column interpretation.} Bronze = SANY fail rate; Silver = SANY
pass rate ($\geq$ silver tier); Gold = TLC pass rate ($\geq$ gold
tier); Diamond = mutation-sensitive pass rate. Bronze and Silver sum to
100\%; Gold and Diamond are subsets of Silver.}
\label{tab:headline_results}
\begin{tabular}{lrrrr}
\toprule
Model & Bronze & Silver & Gold & Diamond \\
\midrule
FormaLLM avg.\ (untuned) & 73.4\% & 26.6\% & 8.6\% & N/A \\
\ours{} SFT-only & 33.3\% & 66.7\% & 13.3\% & 13.3\% \\
\ours{} +DPO (ablation) & 26.7\% & 73.3\% & 23.3\% & 20.0\% \\
\textbf{\ours{} +repair-GRPO} & \textbf{20.0\%} & \textbf{80.0\%} & \textbf{30.0\%} & \textbf{30.0\%} \\
\bottomrule
\end{tabular}
\end{table*}

%% file: tables/tab_per_stage.tex
\begin{table*}[!t]
\centering
\small
\caption{Per-tier results on the held-out 30-problem suite showing
the cumulative production training path. The SFT + DPO row is a
same-base ablation; both it and the final repair-GRPO checkpoint use
the SFT checkpoint as their base, not each other.
$\Delta$(G--D) counts specs that pass Gold but fail Diamond
(mutation-insensitive invariants).}
\label{tab:per_stage}
\begin{tabular}{lrrrr}
\toprule
Stage & SANY $\geq$ silver & Gold & Diamond & $\Delta$(G--D) \\
\midrule
gpt-oss-20b base (0-shot) & 13.3\% & 0\% & 0\% & 0 \\
+ SFT (1 epoch) & 66.7\% & 13.3\% & 13.3\% & 0 \\
SFT + DPO (ablation) & 73.3\% & 23.3\% & 20.0\% & 1 \\
+ repair-GRPO (160 steps) & 80.0\% & 30.0\% & 30.0\% & 0 \\
\bottomrule
\end{tabular}
\end{table*}

%% file: figures/fig_training_progression.tex
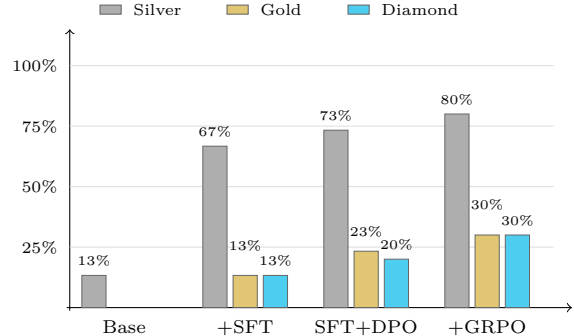
\begin{figure}[!t]
\centering
\begin{tikzpicture}[font=\footnotesize, scale=0.8]
  \draw[->, line width=0.4pt] (-0.05,0) -- (8.3,0);
  \draw[->, line width=0.4pt] (0,-0.05) -- (0,4.6);
  \foreach \y in {25,50,75,100} {
    \draw[gray!25] (0, \y/25) -- (8.0, \y/25);
    \node[anchor=east, font=\tiny] at (-0.05, \y/25) {\y\%};
  }

  \fill[silvercolor!70, draw=black!60, line width=0.2pt] (0.2,0) rectangle (0.6, 13.3/25);
  \node[above, font=\fontsize{5}{6}\selectfont] at (0.4, 13.3/25) {13\%};

  \fill[silvercolor!70, draw=black!60, line width=0.2pt] (2.2,0) rectangle (2.6, 66.7/25);
  \node[above, font=\fontsize{5}{6}\selectfont] at (2.4, 66.7/25) {67\%};
  \fill[goldcolor!70,   draw=black!60, line width=0.2pt] (2.7,0) rectangle (3.1, 13.3/25);
  \node[above, yshift=4pt, font=\fontsize{5}{6}\selectfont] at (2.9, 15.3/25) {13\%};
  \fill[diamondcolor!70,draw=black!60, line width=0.2pt] (3.2,0) rectangle (3.6, 13.3/25);
  \node[above, font=\fontsize{5}{6}\selectfont] at (3.4, 13.3/25) {13\%};

  \fill[silvercolor!70, draw=black!60, line width=0.2pt] (4.2,0) rectangle (4.6, 73.3/25);
  \node[above, font=\fontsize{5}{6}\selectfont] at (4.4, 73.3/25) {73\%};
  \fill[goldcolor!70,   draw=black!60, line width=0.2pt] (4.7,0) rectangle (5.1, 23.3/25);
  \node[above, font=\fontsize{5}{6}\selectfont] at (4.9, 25.3/25) {23\%};
  \fill[diamondcolor!70,draw=black!60, line width=0.2pt] (5.2,0) rectangle (5.6, 20.0/25);
  \node[above, font=\fontsize{5}{6}\selectfont] at (5.4, 20.0/25) {20\%};

  \fill[silvercolor!70, draw=black!60, line width=0.2pt] (6.2,0) rectangle (6.6, 80.0/25);
  \node[above, font=\fontsize{5}{6}\selectfont] at (6.4, 80.0/25) {80\%};
  \fill[goldcolor!70,   draw=black!60, line width=0.2pt] (6.7,0) rectangle (7.1, 30.0/25);
  \node[above, yshift=4pt, font=\fontsize{5}{6}\selectfont] at (6.9, 32.0/25) {30\%};
  \fill[diamondcolor!70,draw=black!60, line width=0.2pt] (7.2,0) rectangle (7.6, 30.0/25);
  \node[above, font=\fontsize{5}{6}\selectfont] at (7.4, 30.0/25) {30\%};

  \node[anchor=north, font=\scriptsize] at (0.9,  -0.05) {Base};
  \node[anchor=north, font=\scriptsize] at (2.9,  -0.05) {+SFT};
  \node[anchor=north, font=\scriptsize] at (4.9,  -0.05) {SFT+DPO};
  \node[anchor=north, font=\scriptsize] at (6.9,  -0.05) {+GRPO};

  \fill[silvercolor!70, draw=black!60] (0.5, 4.85) rectangle (0.85, 5.0);
  \node[anchor=west, font=\tiny] at (0.9, 4.925) {Silver};
  \fill[goldcolor!70,   draw=black!60] (2.6, 4.85) rectangle (2.95, 5.0);
  \node[anchor=west, font=\tiny] at (3.0, 4.925) {Gold};
  \fill[diamondcolor!70,draw=black!60] (4.6, 4.85) rectangle (4.95, 5.0);
  \node[anchor=west, font=\tiny] at (5.0, 4.925) {Diamond};
\end{tikzpicture}
\caption{Pass-rate progression across training stages on the 30-problem
held-out suite. Silver bars show the SANY pass rate; Gold and Diamond
bars show cumulative TLC and mutation-test pass rates respectively.
The DPO checkpoint is a same-base ablation; the production
path runs: base, SFT, repair-GRPO. The single Gold-Diamond
gap at the DPO stage ($\Delta{=}1$) is the only instance where a
tautological invariant is detected across all checkpoints.}
\label{fig:training_progression}
\end{figure}

%% file: tables/tab_domain_breakdown.tex
\begin{table}[!t]
\centering
\scriptsize
\setlength{\tabcolsep}{4pt}
\caption{Per-domain pass-rates on the 30-problem held-out suite at
the final repair-GRPO checkpoint.}
\label{tab:domain_breakdown}
\begin{tabular}{lrrr}
\toprule
Domain & $n$ & Gold & Diamond \\
\midrule
Communication    & 3 & 0 & 0 \\
Concurrency      & 3 & 0 & 0 \\
Consensus        & 3 & 1 & 1 \\
Data structures  & 3 & 1 & 1 \\
Memory / caches  & 3 & 1 & 1 \\
Mutual exclusion & 3 & 2 & 2 \\
Puzzles          & 3 & 0 & 0 \\
Scheduling       & 3 & 2 & 2 \\
Transactions     & 3 & 0 & 0 \\
Workflows        & 3 & 2 & 2 \\
\midrule
\textbf{Total}   & \textbf{30} & \textbf{9} & \textbf{9} \\
\bottomrule
\end{tabular}
\end{table}

%% file: figures/fig_output_example.tex
\begin{figure*}[!t]
\centering
\begin{minipage}[t]{0.47\textwidth}
\begin{tcolorbox}[
  title={ Diamond-tier output},
  colback=gray!5,
  colframe=diamondcolor,
  colbacktitle=diamondcolor,
  coltitle=white,
  fonttitle=\small\bfseries,
  boxrule=0.8pt,
  arc=3pt,
  left=4pt, right=4pt, top=2pt, bottom=2pt
]
\begin{verbatim}
---- MODULE Counter ----
EXTENDS Naturals
VARIABLE n
Init   == n = 0
Next   == n' = (n+1) % 3
TypeOK == n \in 0..2
Spec   == Init /\ [][Next]_n
====
\end{verbatim}
\end{tcolorbox}
\end{minipage}\hfill
\begin{minipage}[t]{0.47\textwidth}
\begin{tcolorbox}[
  title={Common failure (tautology)},
  colback=gray!5,
  colframe=red!60!black,
  colbacktitle=red!60!black,
  coltitle=white,
  fonttitle=\small\bfseries,
  boxrule=0.8pt,
  arc=3pt,
  left=4pt, right=4pt, top=2pt, bottom=2pt
]
\begin{verbatim}
---- MODULE Broken ----
EXTENDS Naturals
VARIABLE v
Init   == v = 0
Next   == v' = v
TypeOK == TRUE
Spec   == Init /\ [][Next]_v
====
\end{verbatim}
\end{tcolorbox}
\end{minipage}
\caption{Diamond-tier output (left) versus a common failure mode (right).
The diamond output's invariant \texttt{TypeOK\,{==}\,n\,\textbackslash{}in\,0..2}
is mutation-sensitive; mutating the range causes TLC to find a counterexample.
The failing output's invariant \texttt{TypeOK\,{==}\,TRUE} passes TLC
vacuously on every reachable state and fails the diamond mutation check.}
\label{fig:output_example}
\end{figure*}

%% file: tables/tab_ablation_tier.tex
\begin{table}[!t]
\centering
\small
\setlength{\tabcolsep}{4pt}
\caption{Effect of the SFT corpus curation cutoff on Stage-1
(SFT-only) Diamond pass-rate on the held-out suite.}
\label{tab:ablation_tier}
\begin{tabular}{lrr}
\toprule
SFT corpus tier-cutoff & rows & Diamond rate \\
\midrule
silver-and-above & 4{,}210 & 6.7\% (2/30) \\
gold-and-above & 1{,}890 & 10.0\% (3/30) \\
\textbf{diamond-only} & \textbf{1{,}053} & \textbf{13.3\% (4/30)} \\
\bottomrule
\end{tabular}
\end{table}

%% file: tables/tab_augmented_inference.tex
\begin{table*}[!t]
\centering
\small
\caption{Augmented inference strategies on the final checkpoint, 30-problem suite.}
\label{tab:augmented_inference}
\begin{tabular}{lrr}
\toprule
Inference strategy & Diamond & Notes \\
\midrule
Greedy ($T=0$, $K=1$) & 9/30 (30\%) & headline \\
Best-of-$K$ ($T=0.5$, $K=8$) & 11/30 (37\%) & +2 wins \\
Greedy~$\cup$~bo-8 union & 13/30 (43\%) & ceiling \\
3-shot TLC-feedback repair & 9/30 (30\%) & no gain \\
\bottomrule
\end{tabular}
\end{table*}

%% file: sections/conclusion.tex
\section{Conclusion}
\label{sec:conclusion}

\ours{} is a trained 20-billion-parameter model
that reaches 30\% at both Gold and Diamond on a held-out 30-problem
\tlaplus{} benchmark, roughly $3.5\times$ the FormaLLM
baseline of 8.6\% TLC model-check. Three contributions support this result.
First, a four-tier validation hierarchy with a mutation-test top tier
blocks always-true invariants from earning credit.
Second, repair-based GRPO outperforms the DPO ablation by 10 percentage
points at Diamond; online feedback from TLC is a stronger training signal
than static preference pairs.
Third, gold and diamond pass-rates coincide at every production checkpoint.
This rules out the always-true invariant reward hack empirically.
\autoref{sec:discussion} documents six instructive failures covering corpus
mislabeling, evaluation variance, GRPO reward collapse, SFT stacking
regression, template memorization and training-target comment leakage.

\BfPara{Limitations} The reward signal is deterministic but not free;
each TLC run takes 1--60 seconds depending on spec complexity, which
limits RL throughput to roughly $K{=}4$ samples per training step on
a two-GPU setup. The 30-problem holdout is small and each problem
swapped in or out shifts the headline by 3.3 percentage points. The
current checkpoint specializes in invariant-style specifications and
remains weak on protocols that require case-analysis or quantifier
manipulation over message types. Diamond-tier mutation testing blocks
tautology-style reward hacks but does not catch subtler failure modes,
such as invariants that are non-trivial yet miss the protocol's
interesting behavior.

\BfPara{Future Work}
A larger holdout of 100--200 problems would shrink per-problem
variance and allow reliable domain-level breakdowns. Pairing TLC with
Apalache's symbolic model-checker~\cite{konnov2019apalache} would
strengthen verification on specs whose state space exceeds TLC's
enumeration budget. A parallel line of work targets TLAPS proof
acceptance as a reward signal; that dataset and training pipeline are
described in a companion paper in preparation.
The current model reuses a narrow specification template for most diamond-tier outputs.
Diverse benchmark problems from message-passing and phase-transition protocols would test whether the training recipe generalizes.
The diamond tier blocks tautological invariants but not semantically weak ones.
Richer mutation operators, such as invariant-strengthening and state-pruning mutations, would close that gap.

\BfPara{Reproducibility}
The final adapter, training code, four-tier curation pipeline,
held-out benchmark with its TLC configurations and the full
training-data manifest will be released publicly on acceptance.
Hosting URLs will be provided in the camera-ready version.

Code: \url{https://github.com/LUC-AI4FM/TLA-Prove}

Dataset: \url{https://github.com/LUC-AI4FM/FormaLLM}